# Authoris: a tool for authority control in the Semantic Web

**Amed Leiva-Mederos, José A. Senso, Sandor Domínguez-Velasco** and
**Pedro Hípola**

Departamento de Informacion y comunicación
Universidad de Granada, Granada, Spain

## Structured abstract

### Purpose

The paper proposes a tool that generates authority files to be integrated with Linked Data by means of learning rules.

AUTHORIS is software developed to enhance authority control and information exchange among bibliographic and non-bibliographic entities.

### Design/methodology/approach

The article analyzes different methods previously developed for authority control as well as IFLA and ALA standards for managing bibliographic records. Semantic Web technologies are also evaluated.

AUTHORIS relies on Drupal and incorporates the protocols of Dublin Core, SIOC, SKOS and FOAF. The tool has also taken into account the obsolescence of MARC and its substitution by FRBR and RDA.

Its effectiveness was evaluated applying a learning test proposed by RDA (2011). Over 80% of the actions were carried out correctly.

### Findings

The use of learning rules and the facilities of Linked Data make it easier for information organizations to reutilize products for authority control and distribute them in a fair and efficient manner.

### Research limitations/implication

The ISAD-G records were the ones presenting most errors. EAD was found to be second in the number of errors produced. The rest of the formats —MARC 21, Dublin Core, FRAD, RDF, OWL, XBRL and FOAF— showed fewer than 20 errors in total.

**Practical implications**

AUTHORIS offers institutions the means of sharing data with a high level of stability, helping to detect records that are duplicated and contributing to lexical disambiguation and data enrichment.

**Originality/value**

The software combines the facilities of Linked Data, the potency of the algorithms for converting bibliographic data, and the precision of learning rules.

# 1.- Introduction

The need to improve interoperability within the World Wide Web gave rise to the development of the Semantic Web, which in turn led to the appearance of many new ways to control and standardize the description of documents, solve problems surrounding diverse indexing systems, and improve the interoperability of records (SKOS, SIOC, Dublin Core, FOAF, etc.). Authority control is a global problem, affecting not only libraries but organizations of all kinds. Publication of authority data on the Web in a heterogeneous or arbitrary way produces inefficiency in information retrieval and creates complications when attributing authority to a given work.

The library community has long been aware of the need for authority control. Evidence of this aim can be traced to the 19th century, when the first Regulations for the production of catalogues, entitled *Rules for a Printed Dictionary Catalogue*, came out in 1876 by Charles Cutter (Cutter, 1986). It was followed almost immediately by *Anglo-American Cataloguing Rules* (AACR), described in works by Fumagalli (Fumagalli, 1887) and Tillett (Tillett, 2004). Both are still used today as control tools in libraries and information agencies all over the world.

Libraries and organisms of international prestige such as the Library of Congress, the Bibliothèque Nationale de France and IFLA unite forces to share data and thereby contribute to authority control. These bodies acknowledge the fact that the information exchange protocols on the Web are insufficient means of controlling authority in the catalogues and systems of library management, since not all countries and organizations can deploy the same level of technological or human resources in their cataloguing efforts, making cooperative cataloguing nearly impossible.

The OCLC, IFLA and the US Library of Congress have fueled initiatives for authority control by sharing the records of various cataloguing agencies. Fruit of this work is t*he Virtual International Authority File* (VIAF), which has meant advances in the construction and generation of authority entries, though it has not reached all the major information institutions at the international level (Bourdon and Zillhardt, 1997). The fact that this project makes it possible to contact information organizations and recognize authority records within an exclusive setting that embraces over 10 National Libraries is, at the same time, a matter that limits the possibilities for outside organizations attempting to access the high quality records generated in these libraries. Developing VIAF records to interact beyond the library framework calls for adopting a more open, interactive, non-exclusive, and ultimately operative viewpoint for authority control. For this reason, we need to take a better look at the processes that actually facilitate collaboration with all organisms producing, consuming or diffusing information. The use of learning rules and the facilities of Linked Data make it easier for information organizations to reutilize products for authority control and distribute them in a fair and efficient manner. Hence, the aim of the present contribution: to propose a tool that generates authority files to be integrated with Linked Data by means of learning rules.

**2.- Material and Methods**

The proposed tool combines the potential of Linked Data with the strengths of the protocols of the Semantic Web, to which we added working rules based on the experience of librarians in creating access points (Tillett, 2004). To test the effectiveness of AUTHORIS we used an efficiency measure, connected to the catalogues of 34 libraries in Spain and the United States. Two variables were used for evaluation: capacity to create records, and quality in record creation. The indicators assessed for each variable are defined in the evaluation section (section 6).

## 3.- Authority control: genesis

Authority control is a matter that has exacted the effort of generations of librarians and cataloguers. The need to uniformly record information on each author included in a catalogue is addressed in work and research stemming from several international organizations. Thanks to the efforts of the IFLA, LITA and ALA, among others, the community of cataloguers has come to adopt standards for generating catalogue entries in a homogeneous manner. The overall goal is to unify the criteria of diverse library consortia and create bibliographic records guided by universally described and accepted cataloguing standards. A brief outline of the development of authority control would include the following landmarks:

- The need for authority control is made explicit, and the Name Authority Cooperative (NACO) comes to light within the US Library of Congress. In Asia, the Hong Kong Chinese Authority Name (HKCAN) is established. This meant recognition of the issue in just two organisms worldwide — far, however, from the syndetic goals set forth by Charles Cutter in the 19th century (Cutter, 1986).

- Lubetzky (Lubetzky, 1969) improves the search and retrieval of authored works in bibliographic records, eliminating the deficiencies that interfered with the retrieval and location of authors in a catalogue.

- Ritvars Bregzis (Bregzis, 1982) creates the ISADN (International Standard Authority Data Number) to overcome difficulties when retrieving bibliographic records with works relative to a given author and with works recorded under a uniform title. The ISADN became an important tool to connect bibliographic records of diverse authors and multiple levels of citation, using fixed numbers for each author or associated work. Soon, the ISADN was key to operations in US libraries.

- The Control Interest Group (ACIG), created under the ALA in 1984, can be seen as a quantum leap in the development of cataloguing activity. Its research into authority control in the United States focused on the use of catalogues in public and specialized libraries to make recommendations for the uniform treatment of authority.

- The guidelines known as *Functional Requirements for Bibliographic Records* (FRBR) and the AACR2 facilitate the use of several tags for information retrieval (Pino, 2004; Danskin, 1996; 1998), among them: **Find**: to locate an entity or entities through attributes and relations; **Identify**: confirms the correspondence between the records searched for and the ones actually located; **Obtain**: facilitates acquisition of an entity or item; and **Navigate**, which helps both the cataloguer and general public to navigate through materials related with their search in the document collection.

- MARC appears on the international scene, along with national formats such as IBERMARC and UKMARC, plus versions like UNIMARC and MARC 21. Thus, bases are established for worldwide cooperation among entities and the interchange of records.

- IFLA proposes the ISAN, International Standard Authority Number, and the codes of the ISO International Standard Text Code family (ISTC) to identify works and expressions, thereby facilitating information exchange within organizations on an international level.

Procedures and standards largely described in the 1960s and the 1990s therefore served as the starting point for the development of authority control in the digital realm.

**4.- Internet, the Semantic Web and VIAF: union of procedures for authority control.**

The Semantic Web, introduced by Berners-Lee, Hendler and Lassila, paved the way for Web interoperability and the existence of formats for information exchange and ontologies (Berners-Lee *et al.*, 2001), elements that bring authority control into an increasingly flexible arena. Below we highlight some elements that, in our view, must be taken into account when building tools for authority control on the Semantic Web.

1. The growth of digital libraries sets the stage for the Z39.50 protocol to enhance interoperability. Thanks to the strong points of this protocol, copy cataloguing is easier, as is the reutilization of bibliographic records.
2. The connections between MARC, MARC 21 and XML are made visible. The relations among these formats for information display and retrieval facilitate mapping between databases, thereby making bibliographic exchange easier for information institutions. The novel appearance of FRBR*,* with its ensuing consolidation, constitutes a turning point for information description, affording flexibility in data processing tasks.
3. IFLA publishes the *Guidelines for Authority Records and References* (GARR). The archives community develops EAD to encode authority metadata using XML, in turn, connected to formats MARC 21 and Dublin Core.
4. The so-called FRANAR (*Functional Requirements and Numbering of Authority Records*) appear on the scene, integrating elements of access and processing in the realm of authority records (Bourdon, 2002). The remarkable aspects of FRANAR stem from its capacity to agglutinate, from a single record, elements specifying information about the author in all its dimensions. FRANAR heightens the potential for information retrieval from bibliographic records by means of the following options: **Search** (for authors or entities), **Identify** (authors or entities), **Control** (creating mechanisms for authority control) and **Relate** (showing resources relative to an item). Moreover, it manages records by: **Process, Sort** and **Display.** Both FRBR and FRANAR are standards that allow for creating authority records with diverse links, facilitating the establishment of different types of relations. In this paper, we use them as models to structure authority records.

Deserving special mention are RDA and VIAF, in view of their importance for the development of our application, AUTHORIS.

RDA (Resource Description and Access) appears in 2005 as a standard for data description and exchange (US RDA Test Coordinating Committee, 2011). It includes the *Functional Requirements for Bibliographic Records* (FRBR) and the *Functional Requirements for Authority Data* (FRAD), which filled the gap of a descriptive cataloguing standard allowing for:

- Quick insertion into the dynamic context of libraries and other producers and users of information.
- Flexible relations between or among entities.
- Greater use and data management in conjunction with digital media.
- More precise description (beyond the possibilities of existing formats) of printed monographs and serial publications.
- Greater ease when using cataloguing metadata in Linked Data operations, aside from ensuring data tagging to facilitate exchanges among bibliographic and non-bibliographic organizations.
- Flexibility, departing from the exclusive focus on Anglo-American rules, meaning metadata can be easily reutilized.

RDA is nourished by new IFLA principles and stands as a noteworthy cataloguing advance toward object-oriented databases. The Web environment called for improving aspects such as the recognition of bibliographic contents, the use of bibliographic data by search engines, and the inclusion of user needs in the processes of describing resources.

The development of VIAF is the result of what we explained earlier in section 3, the genesis of authority control in digital environments. It was carried out by the Library of Congress, Deutsche Nationalbibliothek, Bibliothèque Nationale de France, OCLC and other worldly National Libraries (OCLC, 2003). VIAF pursues the goal of connecting the authority records of diverse libraries in a global information service, with standardization of different names for the same person or organization. According to Boeris, VIAF has become the venue for a vast community of libraries and agencies to reconfigure their bibliographic data so as to better serve users of different languages (Boeris, 2011).

The VIAF initiative is organized by OCLC, in charge of revising and comparing records containing author names and their assignments, as well as the documents existing in national bibliographic records and in the WorldCat. Each VIAF record is generated with information drawn from the comparison of records, and it includes underlying data from authority catalogues and bibliographic catalogues (see figure 1).

Figure 1. VIAF Record

Nonetheless, the VIAF objective cannot be achieved by all libraries worldwide due to the fact that:

1. It does not take into account the cognitive factors of the user, and at times describes matters in "librarian" terminology, opaque for the average Web user.

2. Librarians cannot modify or improve their catalogue entries.

3. Not all organizations around the world use standards and guidelines such as GARR, GSARE, AACR2 or RDA.

4. Cataloguing work is not organized all over the world in such a way that would permit the formation of work groups to attend VIAF in all regions.

5. Even though most countries have National Libraries, these may not be the most appropriate organisms for coordinating VIAF activities.

6. Insufficient qualified human resources for cataloguing and a lack of regulations and work standards translate as frequent errors in the entries of catalogues.

7. Not all the information resources on authority are generated by information organizations, for which reason the work of cataloguing and control is more complicated in the Web environment.

8. VIAF works with a small number of entities and organizations. There is demand for a model of interchange that would be rooted in information resources of a public nature, not necessarily generated from the domain of librarians.

9. There is a need for software that would monitor errors, homogenizing Web searches and authority entries, aside from being able to recognize and group elements under a uniform title. Those existing at present respond to the demands of librarians, not those of other organisms where bibliographic information is commonly used.

In order to implement the VIAF postulates and authority control within the Semantic Web, or in Linked Data, a new dimension for processing information on the Web is needed (Greenberg and Robertson, 2002). To proceed in this direction, the bibliographic records of libraries must be correctly mixed with the protocols of the Semantic Web and similar organizations. In this sense, we can underline the efforts invested by the Library of Congress from 2005 onwards, with their contributions from MARC through XML, MODS (Metadata Object Description Schema) and MADS (Metadata Authority Description Schema).

A further essential element for adapting to the Semantic Web lies in the authority control by information organizations (Qiang, 2004; Taylor, 1999) and, more specifically, their conversion to RDF (Resource Description Framework) (Miles *et al.*, 2005). This is a complex task that sometimes entails huge programming costs, as organizations may have records in diverse bibliographic formats.

Such obstacles do not detract from the semantic wealth of these vocabularies, with potential for constructing linguistic structures that improve the ratio of recall and precision in the retrieval of information from the Web. The lexical structures involved enable diverse users to engage in communication, overcoming linguistic barriers when searching for or retrieving information.

Human and technological resources on the library horizon show great variation. Many libraries, for instance that of the University of North Carolina, devise systems with a high syndetic level; whereas others might not even have generated catalogues in a first level of bibliographic description. Such great differences across the board underscore the significance of even attempting a global system for authority control. Along with the phenomenon of diverse technological capacities, we encounter a proliferation of tools produced through the very development of the Web, by non-librarian organizations (Harper and Tillett, 2007). The appearance of exchange protocols on the Semantic Web (FOAF, SKOS, Dublin Core), their linkage and simplification, are making it possible to generate records manageable for any user or level of technology.

One of the most popular protocols in the context of the Semantic Web is FOAF (Friend of a Friend), developed under the RDF scheme. Based on the principles of authority dictionaries and Who's Who directories, FOAF facilitates author interconnection and interrelations. This format stands as a sound option for representing the data assigned to a person, e. g., name, surname(s), e-mail address, topics of interest, research projects, publications..., patterns that can be applied to any authority record.

## 5.- AUTHORIS
*5.1. Conception and underlying methodology*

AUTHORIS aspires to facilitate the processing of authority data in a standardized fashion, following the principles of Linked Data. Unlike systems such as Virtual Open Access in Agriculture and Aquaculture Repository (VOA3R), AUTHORIS can be used by all sorts of bibliographic agencies, publishing companies, associations or libraries. This software was produced by members of the Department of Information and Communication of the University of Granada, Spain, whose collaborative goal is to develop a platform for information exchange and authority encoding. The tool has taken into account the obsolescence of MARC and its substitution by FRBR and RDA.

The software features multiple functionality, to favor transformation of bibliographic records and to determine uniform entries for corporate authors as well as individual ones. Each function is derived from the facilities of Linked Data, the potency of the algorithms for converting bibliographic data, and the precision of learning rules. AUTHORIS relies on Drupal, a CMS (Content Management System) that works with semantic information, and contains the protocols of Dublin Core, SIOC, SKOS and FOAF.

Drupal is a content management system created by Dries Buytaert in 1999 and developed under GNU license two years later. Creating a website in Drupal consists of combining several "blocks" in order to adapt the site functionality to specific needs. It furthermore provides a Content Management Framework (Byron, Berry and Bondt, 2012). Information is stored in a relational database (it works with MySQL, PostgreSQL, SQLite and others) using PHP programming language.

Drupal permits publication of data in RDF format, or alternative formats such as N-Triples, JSON, XML, RSS 1.0 and Turtle. It handles the URIs of the published RDF data, and accommodates an Endpoint SPARQL for consulting the data. The RDF fields and Namespaces can be personalized.

Hence, it is very flexible software, while featuring a robust security mechanism and sufficient online documentation. A major strength of this CMS resides in the possibility of adding modules. In this case, we opted to generate a new one that would surpass the capacity of the Biblio module, designed to process bibliographic references.

Although there are numerous methodologies for developing information services, in the case of AUTHORIS we opted for the approach devised by Garrido and Tramullas, which highlights the most important aspects for a proper design of digital information (Garrido and Tramullas, 2006). The actions carried out to create the service were:

1. Study the needs of potential users of the service.
2. Develop automata and program text-processing models.
3. Test-run the program.
4. Train staff and create the software documentation.
5. Trial stage of the software.
6. Publication and diffusion of the service.

Below we go over the main features of the system:

- **Authentication:** a user is assigned a role by the system administrator. Each user has specific tasks that include record conversion, automatic cataloguing and information searches.
- **Search for Information:** all users —even those not authenticated by the system— can consult the access points of each record. To this end, they dispose of a system for information search and retrieval based on Semantic Web technologies capable of filtering the records of over 200 information entities (libraries, archives, virtual libraries, etc.). The search for information facilitates data retrieval by author, title, subject, and the combination of Boolean operators for more complex searches, in addition to having a system for document clustering. The search results can be obtained in XML, RDF, FOAF, MARCXML, RDA, FRAD or FRBR format, among others. It is also possible to visualize the information entities where a given record is located, obtain an image of the author's works, and view his/her main collaborators or the bookshops and publishing houses that commercialize the works. The user who searches for authority information can moreover find suggestions about which entries are more complete, or which is recommended for reference.
- **Conversion of files:** converting files is a real strong point of AUTHORIS. A recorded user can import and export files to diverse formats (XML, RDF, FOAF, MARCXML, RDA, FRAD and FRBR). The imported records may be from libraries, publishing companies or other organizations. Within this section one can create new authority entry, by means of learning rules used only in the event that one wishes to produce a uniform title or introduce a new authority. The user merely has to select the nationality of the author (if an individual; or the entry data in the case of a corporative author), and the system automatically assigns the correct entry using the learning rules declared for this purpose. If the system already has the authors introduced by another agency, it will suggest the best entry based on cases previously described and stored. Furthermore, the user can select the output format to export the authority record created. The authority rules are also used to group the uniform titles most used by agencies included in the project. File conversion makes it easier to obtain data in RDF generated by another 200 databases that are not associated with the AUTHORIS project, including DLBP, VIVO, Dbpedia, IEEE, PubChem and Chem2Bio2RDF.

- **Automatic cataloguing:** by means of a Z39.50 client, one can search for and retrieve catalogue records from other libraries in MARC, MARC 21 and MARC XML to complete the cataloguing data in question. This option is proposed for users who do not have RDA-based catalogues and whose national catalogues mainly use MARC 21.

Figure 2. Interface for automatic cataloguing by AUTHORIS

- **Editorial visualization:** thanks to this option, it is possible to extract information referring to a work and its author by means of online consultation of various publishers or libraries.

Figure 3. Data extracted from different publishing companies regarding works by Cervantes

*5.2.- Architecture*

The software is made up of a number of modules for information processing, a system of rules for conversion and authority control, and a Linked Data module. The data sources feeding AUTHORIS are publishing companies, libraries, press agencies, online journals, and databases, whose bibliographic information serves to create authority records similar to the VIAF (see figure 4).

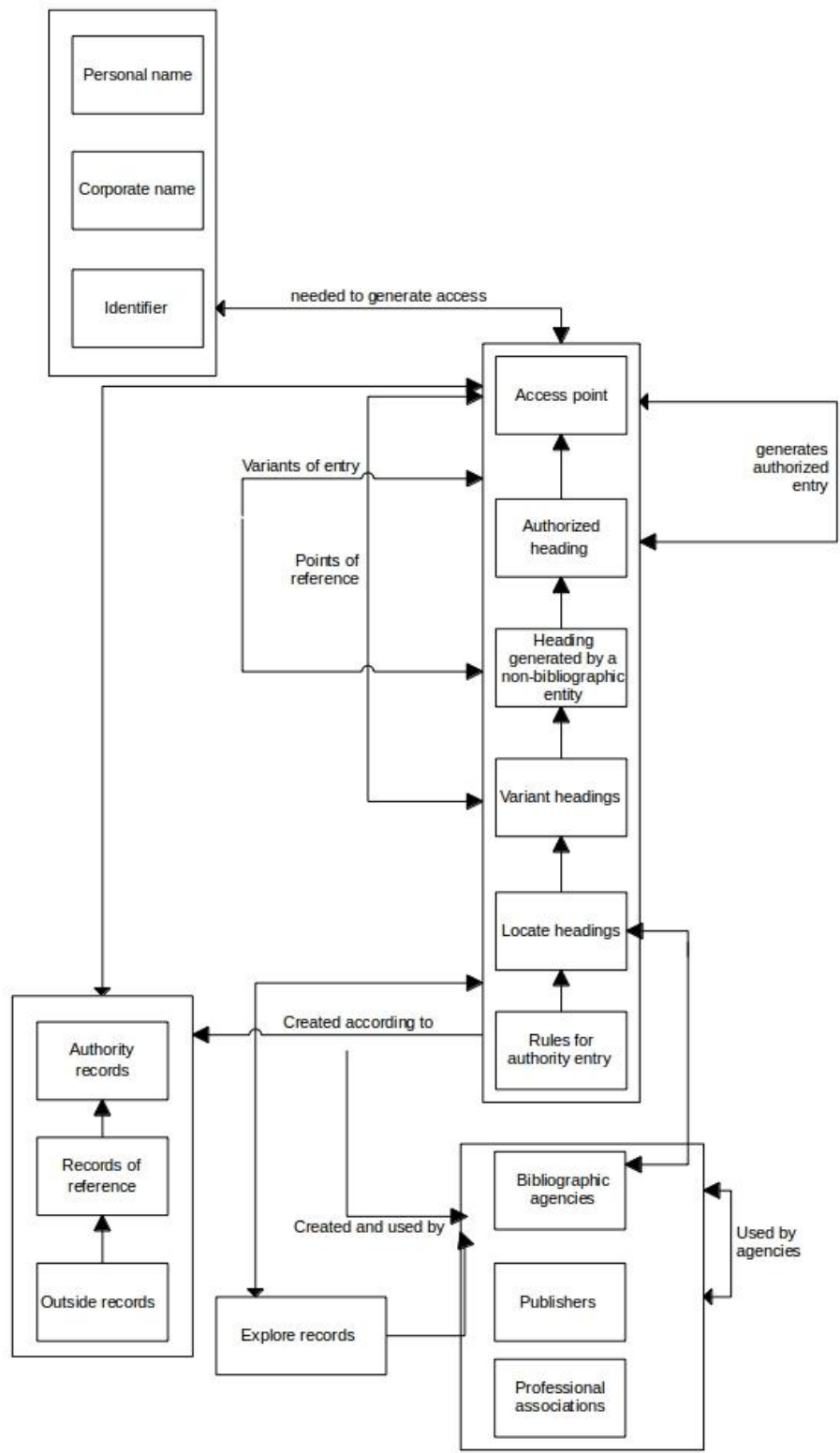

Figure 4. Sketch of exploration and conversion of records in AUTHORIS

The first action undertaken by the system is to access and navigate through authority records generated by bibliographic agencies, publishers and professional associations. Library records are created by means of author entry rules. Those of professional associations and publishers are generated using specific standards adopted by each organization. Once the stage of access and navigation has finished, the system uses rules to generate authorized access points by transforming records into bibliographic standards that may be exchanged with other agencies or organizations. Under AUTHORIS, access entails three elements:

- **Authorized heading:** this is the standard entry generated according to Anglo-American Cataloguing Rules.

- **Heading generated by a non-bibliographic entity:** entries made by other organisms (not libraries, archives or information agencies) that process entries for persons and institutions.

- **Variant headings:** the different forms of subject headings that may be seen in particular cases.

To create an authority record, this software explores and locates records in diverse formats used in other entities, in view of:

- **Authority records:** they present all the information about corporate or individual authors that an information organization may possess. In the authority records, all author entries are standardized. They constitute the foundation of the system and serve to make comparisons between records, to determine which is the best or most complete.

- **Databases:** they contain information on authors and their publications, and include those of DBLP Computer Science Bibliography, Web of Science, Scopus, etc.

- **Virtual publishers and bookshops:** these hold a large portion of the international publishing output worldwide. They take in the titles of works as well as the names of authors.

- **Pages of organizations:** generally speaking, they provide information about persons associated with an organization.

*5.3. Modules*

This tool has four modules, all essential for fulfilling authority control requirements. Processing is channeled through a knowledge base that stores cases referring to any uses of AACR2. Just below we describe the main modules —Linked Data and authority processing— since the other two (information processing and record keeping) are in charge of the tasks of sending and resending records in view of the options chosen.

*Linked Data*

The capabilities of the Linked Data module of AUTHORIS are not fed by the classic proposal of the Semantic Web. This component of the system is able to build links to explore and navigate in the records of any entity by means of SPARQL queries.

The principles guiding Linked Data functions within AUTHORIS are those set forth by Berners-Lee (Berners-Lee, 2006; Méndez *et al.*, 2012):

1. use URIs (uniform resource identifiers*)* to uniquely identify resources;
2. use URI http so that people can access information from the resource;
3. offer information about resources using RDF;
4. include links to other URIs, facilitating links between different data distributed on the Web.

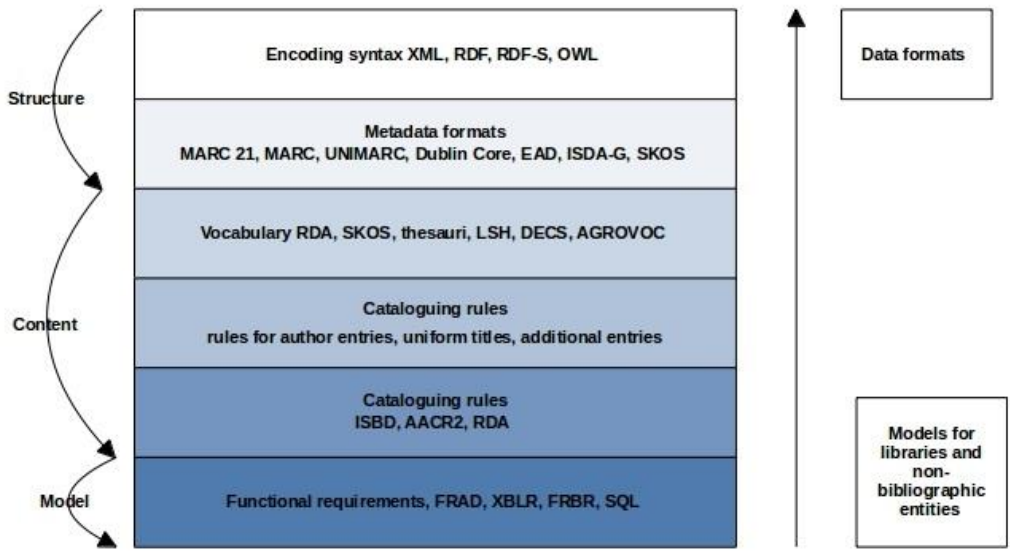

Figure 5. Model of Linked Data of AUTHORIS with format transformation

The Linked Data module is based on the LODE-BD requirements (LOD-enabled Bibliographic Data), a standard providing for the use of bibliographic data (Subirats and Lei Zeng, 2012). With LODE it is possible to use data in any of the formats specified here: SQL, FRAD, XBRL and FRBR.

After activating the data structured in the above formats, the records are converted to ISBD, AACR2 or RDA, depending on the needs of the agency using the software. After transformation, rules to build access are applied, individually, for authors, corporate authors, uniform titles and additional entries. The access rules are complemented by decision-making rules that map the bibliographic records of 32 libraries, capable of determining the most used and most correct entry for each bibliographic record.

After this, within the module of Linked Data it is likewise possible to transform vocabularies or their terminology, converting the records into any of the vocabularies existing in the database of the system, or using vocabularies of reference. The output formats are: XML, RDF, RDFS and OWL.

*Authority rules*

The rules for authority conversion with which we work in this module were derived from a need for systems to be more scalable and professional (Miles *et al.*, 2005), and from daily practices of persons dedicated to information use and management. To transform the data into a bibliographic format, the system converts all sources into XML. This XML file then undergoes transformations into RDF, XML, MARC 21, MARC, UNIMARC or FRBR.

The authority rules were created bearing in mind AACR2 and RDA standards, and regularity in subject headings in libraries all over the world. To formulate these rules, more than 5000 cases (from different organizations) were used. In this way, part of the authority work was "intelligent", carried out automatically, gathering cases where specific conditions were set out for:

- Personal author.
- Corporate author.
- Non-governmental institutions.
- Government institutions.
- International institutions.
- Religious institutions.
- Events.
- Subtitles.
- Parallel title.
- Alternative title.
- Statement of responsibility.
- Shared responsibility.
- Mixed responsibility.

The system of rules for authority control is in charge of standardizing the entries under uniform criteria, which are the same as those used by information organizations worldwide. The rules serve to instrumentalize a learning system that selects an entry by weighing its quality, in view of the following parameters:

- **Rules for the control of individual and corporate authors**: they feed on the Anglo-American Cataloguing Rules and serve to standardize the entries. The system has 300 cases for each rule, and a deciding algorithm that chooses the most complete entry for each heading. Figure 6 illustrates one of the rules employed to process government authority entries.

---

- **Rule 3 Heading for embassies**

- If the heading for an embassy or delegation has no country for which it is accredited, select as valid heading the one held by the country before which it is accredited. Cases:
  - México (embassy) Perú 5 correct
  - México Perú embassy
  - Embassy. México and Perú
  - Selection:
  - If the main entry contains the country, the embassy and the place, it is selected as the most complete entry.
  - **location 1 + embassy 2 + (place 2) make 5**

---

Figure 6. Rules employed to process government authority entries

- **Rules for assigning a uniform title**: these rules function through a comparison of cases. An agent locates the title of the work to be catalogued, and selects cases that repeat a similar title, to then assign the most used title.

- **Rules to assign a standard number to each author:** each author name is standardized using the publisher number or the ISAN, if recorded by some library. Otherwise, the number identifying the author in the publishing company is used. If none of these options is feasible, a code is assigned to record the existence of the author in question.

- **Decision-making rules:** implying the three above rules, they are also referred to as "if, then" rules, in agreement with the weight given to each author record in each particular rule, selecting the most adequate procedure.

Figure 7 depicts the resulting file in RDF format, showing the variants of the author name "Cervantes" located in libraries and publishing companies (110 different forms) based on consultation of the Spanish classic *El ingenioso hidalgo Don Quijote de la Mancha*.

Figure 7. Resulting file in RDF format

## 5.4. State of the application

Both the overall functional framework of AUTHORIS and its respective modules have been in the beta stage since October 2012. Our objective is to have a candidate for the definitive version in the second semester of 2013, and a widely available version by the end of the year. At present, a prototype is in use at the Universidad Central Marta Abreu of Cuba; in this experimental stage the authors have fielded sufficient feedback by users to undertake a comprehensive evaluation of the software.

AUTHORIS is intended to be developed as an Open Source platform, and although it was originally meant to offer a specific service, the final aim is for it to be used by a broad array of groups and libraries. After the trial stage, all the modules configuring AUTHORIS will be available in Google Code under GNU General Public License v 3.0, and processed by means of Git version control.

**6.- Evaluation**

The effectiveness of AUTHORIS was evaluated applying a learning test developed and proposed by RDA (US RDA Test Coordinating Committee, 2011). A total of 16 users participated, divided into the following categories:

- **6 specialists:** graduates of Library and Information Science with 10 years of experience in cataloguing.

- **6 supporting staff:** individuals without training in librarian work.

- **4 other library personnel:** workers with no specialized degrees, in charge of tasks unrelated to bibliographic work.

All of these persons had access to the records of 32 libraries and publishing companies pertaining to Spain and the United States. The personnel selected for evaluation consulted 3000 of the bibliographic records; after obtaining and transforming these data, the participants carried out different actions, including automatic cataloguing, record transformation, navigating and building vocabularies, all intended to transform them into diverse bibliographic formats and generate VIAF-type files.

Then, the quality of the records created was assessed. That is, the effectiveness of the system was evaluated by means of the records generated automatically by AUTHORIS when transforming formats such as MARC, MARC 21, FOAF, Dublin Core, RDA or XBRL into files that could be accessed and reutilized by Linked Data. The elements taken into account in this part of the evaluation were:

a) Omission of the place and date of a conference name when access points were generated for corporate authorities.

b) No inclusion of the terms of relation or designators specifying "issuing body" or "funding body" in the access point main entry when dealing with corporate authors.

c) Omission of ellipses at the beginning of a title of a conference that includes a number in its denomination.

d) Incorrect information about date of recording.

e) No inclusion of the access mode to electronic resources.

f) Omission of the author's family.

g) No declaration of whether an online resource and its printed counterpart are catalogued.

h) Error in the place and date of publication.

i) Omission of the notes on title sources.

j) For sound recordings, incorrect definition of the dates when beginning to record.

k) Incorrect use of uniform titles.

*6.1.- Results*

The ISAD-G records were the ones presenting most errors, above all in the following cases:

- Errors in the placement of data regarding place and date of publication.
- Omission of notes on the title sources.
- Incorrect placement of the dates when recording began, in the case of audio material.

EAD was found to be second in the number of errors produced, with 16 errors of omission of place and date in conferences involving corporate authority. In the transformation to EAD we found 12 records that did not include the terms of relation or designators specifying "issuing body" or "funding body" in the access points of the main entry. 13 EAD records were incomplete and generated confusion, as they did not mention in the description whether an online resource and its printed counterpart were catalogued. These errors can be attributed to the number of fields used by the archive rules and the heterogeneity of cataloguing procedures. The rest of the formats —MARC 21, Dublin Core, FRAD, RDF, OWL, XBRL and FOAF— showed fewer than 20 errors in total, indicating that their transfers are carried out with quality and that they serve for usage with Linked Data.

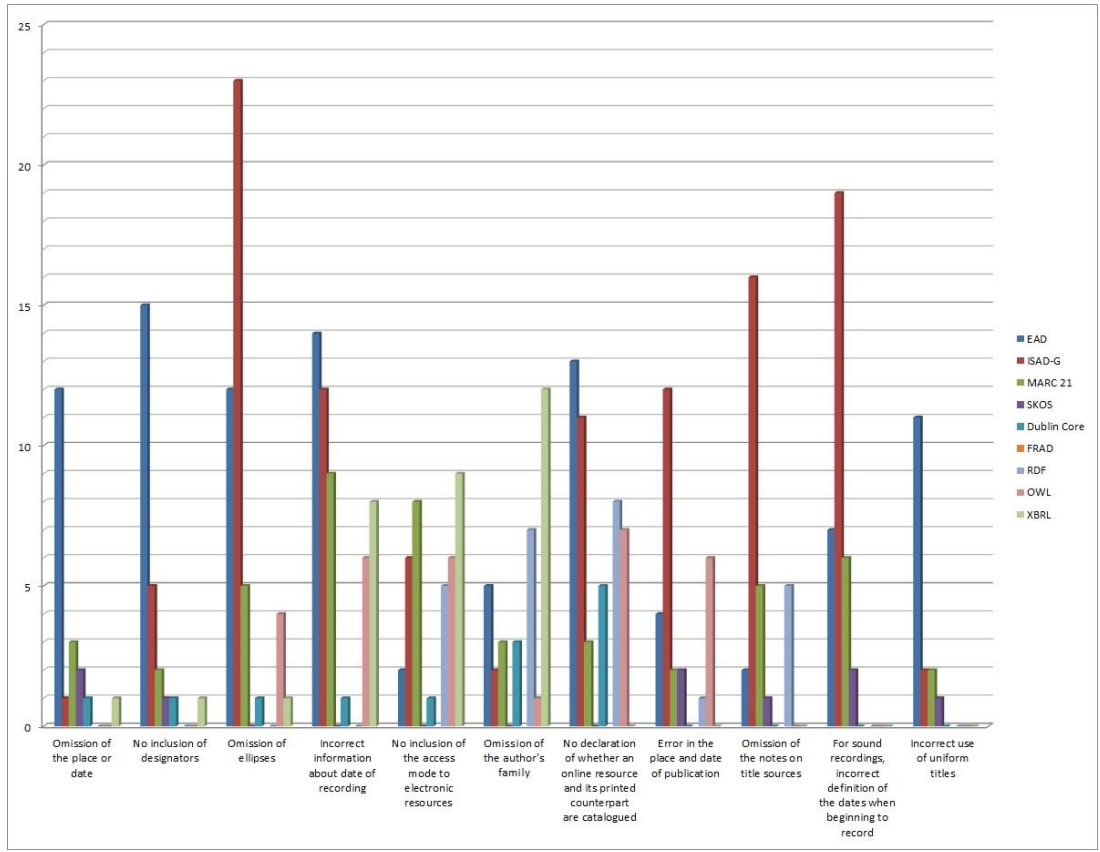

Figure 8. Number of errors produced

### 7.- Conclusions

The development of authority control faces new challenges in the Semantic Web. The need to facilitate interoperability and connection among non-bibliographic and bibliographic entities is one promising area to be implemented by the designers and developers of future cataloguing and authority control systems.

The tool presented in this paper is not meant to be a panacea in this sense. Indeed, the authors hold that such applications will not be needed in the library world of the future, given that integrated library systems will eventually adopt the technologies used by the Linked Data movement as a global reality, not just an extension toward RDA cataloguing.

In the meantime AUTHORIS paves a path toward terrain where future information systems might be oriented: the integration of data and contents. It is our understanding that all efforts should lead to promoting the creation and use of Open Data.

The possibilities lent by the Open Data movement are an aid in accessing remote as well as homogeneous and standardized data, which may be reutilized in other processes. For this reason it is important that libraries open their databases and contribute to the use of formats by non-bibliographic entities.

The model for authority control presented in AUTHORIS is flexible and inclusive. Although certain solutions for international authority control have been put forth previously, progress is slow and tends to be limited to National Libraries or very large ones. Non-library organizations (particularly publishing companies and bookstores) are left out in the cold, despite their capacity to generate vast volumes of authority entries that might be used not only for authority control, but also for information exchange.

AUTHORIS, our proposal in this direction, offers institutions the means of sharing data in a global manner with a high level of stability, moreover helping to detect records that are duplicated, while contributing to lexical disambiguation and data enrichment.

## 8.- References